\documentclass[prl,twocolumn,superscriptaddress,showpacs]{revtex4}

\usepackage{epsfig}

\begin{document}

\title{Phase diagram of silicon from atomistic simulations}

\author{M.~Kaczmarski}
\affiliation{Institut de Ci\`{e}ncia de Materials 
de Barcelona (ICMAB--CSIC), 
Campus de Bellaterra, 08193 Barcelona,
Spain}
\affiliation{Institute of Physics, University of Silesia, Uniwersytecka 4,
40-007 Katowice, Poland}
\author{O.~N.~Bedoya-Mart\'{\i}nez}
\affiliation{Institut de Ci\`{e}ncia de Materials
de Barcelona (ICMAB--CSIC),
Campus de Bellaterra, 08193 Barcelona,
Spain} 
\author{E.~R.~Hern\'{a}ndez}
\thanks{To whom correspondence should be addressed; email: ehe@icmab.es}
\affiliation{Institut de Ci\`{e}ncia de Materials 
de Barcelona (ICMAB--CSIC), 
Campus de Bellaterra, 08193 Barcelona,
Spain} 

\date{\today}

\begin{abstract}
In this letter we present a calculation of the 
temperature-pressure phase diagram of Si in a range of pressures 
covering from $-5$ to 20~GPa and temperatures up to the melting point.
The phase boundaries and triple points between the diamond, liquid, 
$\beta$-Sn and $\mbox{Si}_{34}$ clathrate phases are reported.
We have employed efficient simulation techniques to calculate free
energies and to numerically integrate the Clausius-Clapeyron 
equation, combined with a tight binding model capable of an
accuracy comparable to that of first-principles methods. The resulting
phase diagram agrees well with the available experimental data.
\end{abstract}


\maketitle

Temperature-pressure phase diagrams charter the regions 
of stability of the different allotropes of a material.
The confection of phase diagrams has been a long standing objective of 
experimental physics, chemistry and materials science. However, to date,
the phase diagrams of most materials remain relatively unknown beyond
the domain of normal conditions, because of the technical challenge of 
performing accurate phase behavior studies in conditions of extreme
temperatures and/or pressures. Reliable first principles electronic 
structure calculations~\cite{martin:book} have the potential to be of great
assistance in this problem, and indeed they have proved their value with 
impressive demonstrations of their capabilities, such as the calculation
of the melting curve of iron down to the pressure regime of the Earth's
core~\cite{alfe}, that of aluminum~\cite{vocadlo:alfe}, or 
or that of hydrogen in a similar range of
pressures~\cite{galli}. But such calculations, which employ either
free energy evaluation techniques like thermodynamic integration, or 
directly address phase coexistence by explicitly simulating the interface,
are computationally demanding, and by no means routine. The 
two-phase method, in particular, requires large simulation cells where the 
two phases can be monitored in coexistence, and it is only directly applicable
to solid-liquid equilibria. Nevertheless, in recent years several simulation
techniques have been developed which now make free energy 
calculations~\cite{watanabe:reinhardt,koning1}
and phase boundary determination~\cite{koning2} much more accessible.
In this letter we demonstrate the potential of these
novel techniques by using them to obtain, entirely from atomistic simulations,
the phase diagram of Si in a wide range of temperatures and pressures.

In spite of being one of the most extensively studied materials, 
the phase diagram of Si is not accurately known.
As many as eleven phases other than diamond have been 
identified at high pressures~\cite{silicon}, with at least six of
them being thermodynamically stable in some temperature-pressure domain. It is
now also clearly established~\cite{mcmillan} that other phases become 
stable at negative pressures: the so called clathrate phases~\cite{clathrate}, 
of which $\mbox{Si}_{34}$ (also known as $\mbox{Si}_{136}$) has interesting 
semiconducting properties~\cite{blase} of its own. In Si, as in many 
other materials, computer simulations have been a great aid in 
identifying, and sometimes even predicting~\cite{needs:martin,chang:cohen} 
the occurrence of certain phases, but they have not been much used 
to help establishing the limits of stability of the different phases 
except at zero temperature~\cite{needs:mujica} or 
zero pressure~\cite{sugino:car,alfe:gillan}.
In this letter we calculate the phase diagram
of Si entirely from atomistic simulations, for temperatures 
ranging between 0 and 1700~K, and for pressures in the range $-5$ to 20~GPa.
We have considered four different phases in this temperature-pressure
region, namely the diamond structure (Si {\bf I}), the $\beta$-Sn phase 
({\bf II}), $\mbox{Si}_{34}$ ({\bf C}) and the liquid phase ({\bf L}),
and provide the five corresponding coexistence curves between these,
as well as estimates for the location of the two triple points to be 
found in this area of the phase diagram.

In a recent study~\cite{kaczmarski}, we have shown that certain
tight binding~\cite{tbreview} models, such as those of 
Kwon {\em et al.\/}~\cite{kwon} and Lenosky {\em et al.\/}~\cite{lenosky},
are capable of providing very accurate descriptions not only of the structural
properties of Si, but more importantly of the thermal properties too. 
In fact, these two models predict a melting temperature at zero pressure
which is in better agreement with the experimental value than that
provided by first principles calculations~\cite{sugino:car,alfe:gillan}.
The deciding factor between the Kwon and Lenosky models is that only
the latter correctly predicts a pressure-induced transition from the {\bf I}
to {\bf II} phase~\cite{lenosky}. All calculations reported below
were carried out using supercells containing 128 Si atoms, except
in the case of Si~{\bf C}, where a supercell of 136 atoms was imposed by
the structure. Four special k-points~\cite{monkhorst:pack} were used
to sample the Brillouin zone and provided a sufficient degree of 
convergence even for the metallic phases ({\bf II} and {\bf L}).
All calculations were performed with the {\sc Trocadero} code~\cite{trocadero}. 


\begin{table}[tbh]
\begin{center}
\begin{tabular}{cccc}
\hline\hline
Phase boundary & Pressure & Temperature & dT/dP \\
         & (GPa)    & (K)         & ($\mbox{K GPa}^{-1}$) \\
\hline
 {\bf I--L} & 0 &  $1551 \pm 66$ & $-16$ \\
            &   &  ($1687^a$) & ($-33^a$) \\
 {\bf II--L} & 10 & $1230\pm 25$ & 40 \\ 
 {\bf C--L} & 0 & $1424\pm 57$ & $-73$ \\
            &   & ($1473^b$) & \\
\hline
{\bf I--II} & $15.5 \pm 0.4$ & 500 & $-200$ \\
            & ($10.4-12.4^c$) & ($573^c$) & \\
{\bf I--C} & $-2.47 \pm 0.03$ & 500 & $\approx 10^4$ \\
\hline\hline
\end{tabular}
\end{center}
{\scriptsize $^a$ Reference \onlinecite{silicon},
$^b$ reference \onlinecite{mcmillan},
$^c$ reference \onlinecite{voronin}.} 
\caption{Coexistence points obtained between the different phases considered
in this study. Experimental data is given in parenthesis when available.
These points were later used as starting initial conditions
for dynamical Clausius-Clapeyron integration. The local slope of the 
corresponding coexistence line is also given.} 
\label{table:coexistencepoints}
\end{table}

Let us now briefly describe the procedure adopted for determining the 
phase diagram. Firstly, for all pairs of phases for which a phase boundary
is sought, a coexistence point along the boundary must be found, i.e. a
temperature and pressure for which the Gibbs free energies of the two 
phases in question are equal. If the coexistence line has a small (in 
absolute value) pressure derivative, in order to locate
the coexistence point it is better to fix the pressure of both phases at some 
convenient value, $P$, and to calculate the Gibbs free energy of each
phase in a temperature interval bracketing the coexistence temperature, 
$T_c$, at fixed pressure. This can be done employing the 
{\em reversible scaling\/} technique of de Koning 
{\em et al.\/}~\cite{koning1}. If, on the other hand, the phase boundary
is expected to have a large (in absolute value) pressure derivative, as
is common in solid-solid coexistence lines, it is more convenient to 
fix the temperature and to monitor the Gibbs free energy of each phase as
a function of pressure, which can be done with the {\em adiabatic 
switching\/} technique~\cite{watanabe:reinhardt}.
Once a coexistence point has been thus located, the rest of the phase boundary
is obtained by solving numerically the Clausius-Clapeyron equation: 
\begin{eqnarray}
    \frac{d T_c}{d P} = T_c \frac{\Delta V}{\Delta H},
    \label{eq:clausius-clapeyron}
\end{eqnarray}
where $T_c$ is the coexistence temperature at pressure $P$, and $\Delta V$
and $\Delta H$ are the difference of volumes and enthalpies
of the two phases, respectively. This was done using the dynamical
scheme of de Koning and coworkers~\cite{koning2}. All the above 
techniques require to simulate the system under isothermal-isobaric
conditions, and this was done as described by Hern\'{a}ndez~\cite{hernandez}.
Following the scheme outlined above, 
we proceeded to locate initial coexistence points along the
phase boundaries between the diamond--liquid~({\bf I--L}), 
$\beta$-Sn--liquid~({\bf II--L}), clathrate--liquid~({\bf C--L}), 
diamond--clathrate~({\bf I--C}) and diamond--$\beta$-Sn~({\bf I--II})
phases. Further calculations were carried out at the found
coexistence points in order to quantify the errors incurred in our 
estimations of those conditions. For 
coexistence temperatures at fixed pressure the error can be estimated
from $\delta T_c \approx \delta G/\Delta S$~\cite{deWijs}, where $\delta G$ 
is the error in the Gibbs free energy difference, and $\Delta S$ is 
the entropy change. For coexistence pressures at fixed 
temperature, the error can be estimated from 
$\delta P \approx \delta G/\Delta V$, where $\Delta V$ is the 
change of volume. As well as 
error estimates, these calculations allowed us to
obtain the values of the pressure derivatives of each phase boundary at 
the coexistence point from Eq.~(\ref{eq:clausius-clapeyron}).
Table~(\ref{table:coexistencepoints}) lists the locations of the different
coexistence points, together with their
estimated errors and the local pressure derivatives of the corresponding
coexistence lines.

Our calculated zero-pressure melting point for Si-{\bf I} (1551~K) is in 
reasonably good agreement with the accepted experimental 
value of 1687~K~\cite{silicon}. Although the difference between these values 
may appear to be large, we note that density functional theory~(DFT) 
calculations using the local density approximation~(LDA)
for the exchange-correlation energy predict values in the 
range 1300-1350~K~\cite{sugino:car,alfe:gillan}, and $1492\pm 50$~K when 
a generalized-gradient approximation~(GGA) 
is used instead~\cite{alfe:gillan}. Comparison with our own previous
estimation using the same model~\cite{kaczmarski} gives a difference of
30~K, within our error bars for the melting temperature, although we regard
the present value as more accurate. In agreement with experiments and 
with DFT calculations we obtain a negative slope for the phase boundary.
The melting of (metastable) Si-{\bf II}
at 10~GPa is found to occur at 1230~K. This temperature is already higher
than the experimental estimate of the {\bf I--II--L} triple
point temperature~\cite{voronin}, but is consistent with our own prediction
for this triple point (see below). Like the {\bf I} phase, {\bf C}
has a melting line with a negative slope. For this phase we predict a 
melting temperature at zero pressure of 1424~K, which is only 50~K below
the experimentally measured value at this pressure~\cite{mcmillan}, and
also in good agreement with the calculated value of 
Wilson \&\ McMillan~\cite{wilson:mcmillan}.
As for the coexistence pressure between Si-{\bf I} and Si-{\bf II} at 500~K,
15.5~GPa, it is larger than the experimental values, 10.4-12.4~GPa at
573~K~\cite{voronin}, but we note again that DFT-LDA calculations predict
a value of 8~GPa~\cite{needs:mujica} at zero temperature, while recent 
quantum Monte Carlo 
calculations place it at 16.5~GPa~\cite{alfe:qmc}. At the same temperature, 
the {\bf I--C} coexistence point is found at $-2.47$~GPa, in good agreement
with experimental measurements~\cite{mcmillan} and with both empirical
potential simulations~\cite{wilson:mcmillan} and first principles 
calculations~\cite{ramachandran}.

\begin{figure}[tbh]
\begin{center}
\epsfxsize=8cm
\epsffile{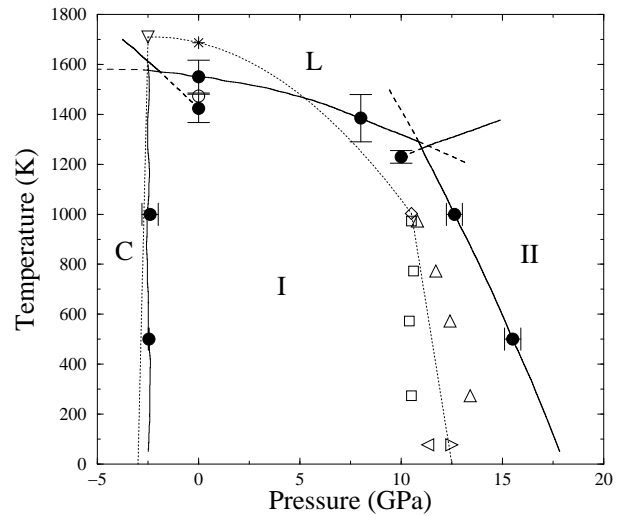}
\end{center}
\caption{Silicon phase diagram. The continuous and dashed lines indicate 
the calculated phase diagram. Dashed curves indicate phase 
boundaries in regions where the separated phases are metastable, 
while continuous black curves 
separate thermodynamically stable phases; uncertainty bounds estimated at
specific points of the phase diagram (marked by filled circles) are 
provided by the error bars. For comparison purposes, a schematic 
phase diagram summarizing the experimental data is shown with dotted lines, and
experimental data at specific temperatures and pressures 
is shown in the form of empty symbols.
The asterisk corresponds to the zero-pressure
melting point of phase {\bf I}, 1687~K~\cite{silicon}; the circle is the 
zero-pressure melting point of the (metastable) {\bf C} phase, at 
1473~K~\cite{mcmillan}; the diamond
is the {\bf I--II--L} triple point, with estimated coordinates 
of $1003\pm20$~K and $10.5\pm0.2$~GPa~\cite{voronin};
empty squares and triangles indicate the pressures at which the 
{\bf II} phase was first observed and where the {\bf I} phase
ceased to be detected, respectively, in the experiments of 
Voronin {\em et al.\/}~\cite{voronin}; left and right pointing triangles
give the same information as obtained by Hu {\em et al.\/}~\cite{hu};
finally, the downward pointing triangle is
the estimated {\bf I--C--L} triple point, at 
1710~K and $-2.5$~GPa~\cite{mcmillan}.}
\label{fig:phase-diagram}
\end{figure}

Taking as starting conditions the coexistence points thus located, we then
proceeded to run dynamical Clausius-Clapeyron~\cite{koning2}
integration calculations, thereby obtaining the sought phase boundaries.
Fig.~(\ref{fig:phase-diagram}) shows our calculated phase diagram for Si,
and constitutes the central result of this work. For comparison,
Fig.~(\ref{fig:phase-diagram}) also shows a summary of 
recent experimental observations from Voronin {\em et al.\/}~\cite{voronin},
Hu {\em et al.\/}~\cite{hu} and McMillan~\cite{mcmillan}.
As can be seen, the calculated phase
diagram captures all the main features of the experimental data
with surprising fidelity. Nevertheless, there are differences in the details
which are mostly attributable to minor shortcomings in the Lenosky
tight-binding model.

The {\bf I--L} coexistence line has, as previously mentioned, a negative 
pressure
derivative, which increases slightly toward larger negative values as the
pressure is raised. This is in agreement with experimental observations
by Voronin {\em et al.\/}~\cite{voronin} that suggest this behavior, although
the pressure derivative predicted by the Lenosky model for this phase boundary
is too small. An independent error estimate of the melting line away from
the starting point of the dynamical Clausius-Clapeyron integration calculation
was obtained at a point of coordinates $T=1385$~K and $P=8$~GPa. We found an 
uncertainty of 95~K, which is not significantly worse than that of the 
zero-pressure melting point. The {\bf I--L} melting line meets the 
{\bf I--II} phase boundary 
at a temperature $T \approx 1290$~K and $P \approx 10.9$~GPa, according to our 
results. The experimental coordinates of this triple point are not 
accurately known, though a recent estimate by Voronin 
{\em et al.\/}~\cite{voronin} puts it at $T = 1003\pm 20$~K and
$P = 10.5\pm 0.2$~GPa. Compared with this best experimental estimate, our
triple point temperature is somewhat too high (by nearly 300~K), consistent
with the fact that $d T_c/dP$ for the {\bf I--L} melting line is too small in
absolute value. However, we emphasize that the estimate of Voronin
{\em et al.\/} is a lower bound; if one assumes that the {\bf I--L} value
of $dT/dP$ remains constant and equal to its zero pressure value, then at
10.5~GPa the triple point temperature should be 1340~K, which is much closer
to our figure. Thus, it is very likely that our triple point temperature 
and that of Voronin {\em et al.\/} provide upper and lower bounds respectively
for the true value.
The agreement in the value of the pressure coordinate is
much better (in fact, within our error estimate for the {\bf I--II} 
phase boundary), but must be understood as somewhat fortuitous, 
resulting from error cancellation between a {\bf I--II} 
coexistence pressure at 500~K
which is slightly too high, and a value of $d T_c/dP$ for the {\bf I--II} phase 
boundary which is most likely smaller than the experimental one. Nevertheless,
it should be pointed out that, to our knowledge, this is the first
prediction from atomistic simulations of the location of the 
of the {\bf I--II--L} triple point of Si.
We have also calculated the {\bf II--L} phase boundary, starting from a 
temperature
and pressure where both the {\bf II} and {\bf L} phases are metastable.
Indeed, the {\bf II--L} coexistence line crosses the {\bf I--II} and {\bf I--L} 
boundaries very
close to the point where the {\bf I--L} and {\bf I--II} boundaries cross, 
a good indication of the internal consistency of our calculations.

The {\bf I--II} phase boundary has, as expected, a large pressure derivative,
of negative sign. Experimental observations~\cite{voronin} seem to agree with 
this finding, although data is only available at large temperature intervals,
and it is not possible at present to compare with an experimental value
of the pressure derivative. In any case, it can be concluded that, 
as it happened with the {\bf I--L} melting curve, the pressure 
derivative of the {\bf I--II} coexistence line
is probably not as large as the experimental one. This can be seen from the
fact that both coexistence pressures at $T = 500$~K and in the
limit $T\rightarrow 0$~K are several GPa higher than the experimental 
coexistence pressures at similar temperatures, while our pressure 
coordinate for the {\bf I--II--L} triple point agrees quite
well with the experimental value. It is worth noting that the
coexistence pressure at $T\rightarrow 0$~K deduced from the calculated 
{\bf I--II} phase boundary (18.1~GPa) is very close to the value obtained by 
calculating the enthalpies of both phases at $T=0$~K (18.5~GPa), which 
serves as another internal consistency check for our calculations. To further
validate our results, we performed an error estimation of the coexistence
pressure at 1000~K. Our results indicate that the coexistence pressure
predicted by the phase boundary at that temperature (12.6~GPa) is accurate
to within 0.4~GPa.  

The {\bf I--C\/} phase boundary we have obtained is nearly a vertical line at
$P\approx -2.5$~GPa. At 500~K our calculated coexistence point occurred
at a pressure of $-2.47\pm 0.02$~GPa 
[see Table~(\ref{table:coexistencepoints})]. At 1000~K, we calculated a second
coexistence point to double check the phase boundary calculation, obtaining
a pressure of $-2.4\pm 0.4$~GPa, to be compared with a value of $-2.5$~GPa
according to our calculated phase boundary. The pressure derivative of this 
phase boundary is so large ($\approx 10^4$~K/GPa in absolute value) as to make
it virtually impossible to predict its sign with any accuracy. This is a
manifestation of the fact that at coexistence, not only the Gibbs free energies
of the two phases are the same, but also their enthalpies are nearly 
equal. This in turn implies that the entropies of the {\bf I} and {\bf C}
phases are very similar at coexistence conditions. The {\bf I--C\/} and 
{\bf I--L} boundaries
cross at $T\approx 1576$~K and $P\approx -2.48$~GPa, marking within our
error bars the coordinates of the {\bf I--C--L} triple point.
Wilson \&\ McMillan~\cite{wilson:mcmillan} have estimated the location of
this triple point to be 1750~K and $-1.5$~GPa, from two-phase coexistence 
calculations employing the Stillinger-Weber potential~\cite{stillinger:weber}.
Experimental estimates~\cite{mcmillan} suggest that the triple point may
actually be closer to 1710~K and $-2.5$~GPa. Again, we note that our predicted
temperature coordinate is too low by about 100-150~K, consistent with our
underestimation of the {\bf I} melting temperature at zero pressure, while the 
pressure coordinate is closer to the experimental estimation.
The {\bf C--L} melting line is almost straight, with a negative slope of
$-73$~K/GPa; it crosses the {\bf I--L} melting line slightly to the 
right of the point where the latter is crossed by the {\bf I--C} boundary,
but their separation is within our estimated error bars.

Thus it is seen that, in spite of its simplicity and semi-empirical nature,
the Lenosky model provides a fairly good description of the phase diagram of
Si, being probably as accurate as could be expected of first principles 
calculations. We have obtained five phase boundaries and two triple points
between four phases of the silicon phase diagram, in reasonable agreement 
with the known experimental
data. The simulation techniques employed in this study to calculate 
free energies and to obtain coexistence curves are straight forward and 
efficient,
and can equally well be used in combination with first principles methods.
It can be concluded, then, that the combination of techniques
used here brings about the possibility of obtaining entire phase diagrams
of complex materials completely {\em ab initio\/}.

The use of computer facilities at $C^4$ (Centre de
Computaci\`{o} i Comunicaci\`{o} de Catalunya), CESGA (CEntro de 
Supercomputaci\'{o}n de GAlicia), ICM (Interdisciplinary Center for 
Mathematical and Computer Modeling, Warsaw University), and TASK
(Academic Computer Center, Gdansk) is gratefully acknowledged. This work has
been funded by the Spanish Ministry of Science and Education (MEC) under
project BFM2003-03372-C03.


\begin{thebibliography}{99}

\bibitem{martin:book} R.~M.~Martin, {\em Electronic structure: basic theory
and practical methods\/} (Cambridge University Press, Cambridge 2004).

\bibitem{alfe} D.~Alf\`{e}, M.~J.~Gillan and G.~D.~Price, 
Nature {\bf 401} 462 (1999).

\bibitem{vocadlo:alfe} L.~Vo\v{c}adlo and D.~Alf\`{e}, 
Phys. Rev. B {\bf 65}, 214105 (2002).

\bibitem{galli} S.~A.~Bonev, E.~Schwegler, T.~Ogitsu and G.~Galli, 
Nature {\bf 431}, 669 (2004).

\bibitem{watanabe:reinhardt} M.~Watanabe and W.~P.~Reinhardt, 
Phys. Rev. Lett. {\bf 65}, 3301 (1990).

\bibitem{koning1} M.~de~Koning, A.~Antonelli and S.~Yip, 
Phys. Rev. Lett. {\bf 83}, 3973 (1999).

\bibitem{koning2} M.~de~Koning, A.~Antonelli and S.~Yip,
J. Chem. Phys. {\bf 115}, 11025 (2001).

\bibitem{silicon} Hull,~R. (Ed.) {\em Properties of Crystalline Silicon\/}
(Inspec, London 1999); M.~I.~McMahon and R.~J.~Nelmes, Phys. Rev. B {\bf 47},
8337 (1993); M.~I.~McMahon, {\em et al.\/}, 
Phys. Rev. B {\bf 50}, 793 (1994); J.~Z.~Hu and I.~L.~Spain, Solid State
Comm. {\bf 51}, 263 (1984); H.~Olijnyk {\em et al.\/},
Phys. Lett. A {\bf 703}, 137 (1984); J.~Crain {\em et al.\/}, Phys. Rev. B
{\bf 50}, 13043 (1994); R.~O.~Piltz {\em et al.\/} Phys. Rev. B {\bf 52},
4072 (1995).

\bibitem{mcmillan} P.~F.~McMillan,
Nature Materials, {\bf 1}, 19 (2002).

\bibitem{clathrate} Clathrates were first synthesized in the 1960s, see
C.~Cros {\em et al.\/}, C. R. Acad. Sci. Paris, {\bf 260\/}, 4764 (1965);
J.~S.~Kasper {\em et al.\/}, Science {\bf 150}, 11713 (1965).

\bibitem{blase} X.~Blase, Phys. Rev. B {\bf 67}, 035211 (2003).

\bibitem{needs:martin} R.~J.~Needs and R.~M.~Martin,
Phys. Rev. B {\bf 30} 5390 (1984).

\bibitem{chang:cohen} K.~J.~Chang and M.~L.~Cohen,
Phys. Rev. B {\bf 31} 7819 (1985).

\bibitem{needs:mujica} R.~J.~Needs and A.~Mujica,
Phys. Rev. B {\bf 51\/}, 9652 (1995).

\bibitem{sugino:car} O.~Sugino and R.~Car,
Phys. Rev. Lett. {\bf 74}, 1823 (1995).

\bibitem{alfe:gillan} D.~Alf\`{e} and M.~J.~Gillan,
Phys. Rev. B {\bf 68}, 205212 (2003).



\bibitem{kaczmarski} M.~Kaczmarski, R.~Rurali,~R. and E.~R.~Hern\'{a}ndez,
Phys. Rev. B {\bf 69}, 214105 (2004).

\bibitem{tbreview} C.~M.~Goringe, D.~R.~Bowler and E.~R.~Hern\'{a}ndez,
Rep. Prog. Phys. {\bf 60}, 1447 (1997).

\bibitem{kwon} I.~Kwon, R.~Biswas, C.~Z.~Wang, K.~M.~Ho and
C.~M.~Soukoulis, Phys. Rev. B {\bf 49}, 7242 (1994).

\bibitem{lenosky} T.~J.~Lenosky, {\em et al.\/}, 
Phys. Rev. B {\bf 55}, 1528 (1997).

\bibitem{monkhorst:pack} H.~J.~Monkhorst and J.~D.~Pack,
Phys. Rev. B {\bf 13}, 5188 (1976).

\bibitem{trocadero} R.~Rurali and E.~R.~Hern\'{a}ndez,
Comput. Mat. Sci. {\bf 28} 85 (2003).

\bibitem{hernandez} E.~R.~Hern\'{a}ndez,
J. Chem.  Phys. {\bf 115}, 10282 (2001).

\bibitem{deWijs} G.~A.~de~Wijs, G.~Kresse and M.~J.~Gillan,
Phys. Rev. B {\bf 57}, 8223 (1998).

\bibitem{voronin} G.~A.~Voronin, C.~Pantea, T.~W.~Zerda, L.~Wang
and Y.~Zhao, Phys. Rev. B. {\bf 68}, 020102 (2003).

\bibitem{ramachandran} Ramachandran,~G.~K. {\em et al.\/}, 
J. Phys.: Condens. Matter {\bf 12}, 4013 (2000); A.~San~Miguel
{\em et al.\/}, Phys. Rev. Lett. {\bf 83}, 5290 (1999);
A.~San~Miguel {\em et al.\/}, Phys. Rev. B {\bf 65}, 054109 (2002).

\bibitem{wilson:mcmillan} M.~Wilson and P.~F.~McMillan,
Phys. Rev. Lett. {\bf 90}, 135703 (2003).

\bibitem{alfe:qmc} D.~Alf\`{e}, M.~J.~Gillan, M.~D.~Towler and
R.~J.~Needs, Phys. Rev. B {\bf 70}, 214102 (2004).

\bibitem{stillinger:weber} F.~H.~Stillinger and T.~A.~Weber,
Phys. Rev. B {\bf 31}, 5262 (1985).

\bibitem{hu} J.~Z.~Hu, L.~D.~Merkle, C.~S.~Menoni and I.~L.~Spain,
Phys. Rev. B {\bf 34}, 4679 (1986).

\end{thebibliography}
\end{document}